\def\re#1{(\ref{#1})}
\def\beq{\begin{equation}}
\def\eeq{\end{equation}}
\def\beeq{\begin{eqnarray}}
\def\beeqn{\begin{eqnarray*}}
\def\eeeq{\end{eqnarray}}
\def\eeeqn{\end{eqnarray*}}
\def\g{\gamma}                  
\def\de{\delta}                 \def\D{\Delta}
\def\e{\varepsilon}

\def\l{\lambda}                 
\def\m{\mu}
\def\n{\nu}

\def\r{\rho}
\def\s{\sigma}                  
\def\th{\theta}

\newcommand{\DD}{{\cal D}}

\newcommand{\OO}{{\cal O}}

\newcommand{\WW}{{\cal W}}

\newcommand{\lp}{\left(}
\newcommand{\rp}{\right)}
\renewcommand{\lq}{\left[}
\renewcommand{\rq}{\right]}

\newcommand{\no}{\nonumber}
\newcommand{\ph}{\phantom} 

\def\tr{\,\mbox{Tr}\,}
\def\frac#1#2{ {{#1} \over {#2} }}

\def\half{\mbox{\small $\frac{1}{2}$}}
\def\p{\partial}
\def\ie{\hbox{\it i.e.}{ }}

\def\bom#1{\mbox{\boldmath$#1$}}

\newcommand{\unity}{1\kern-.65mm \mbox{\form l}}
\newcommand{\ks}{\mbox{\form l}\kern-.6mm \mbox{\form K}}
\newcommand{\A}{A \raise0.5mm\hbox{\kern-1.8mm /}}
\def\pmb#1{\leavevmode\setbox0=\hbox{$#1$}\kern-.025em\copy0\kern-\wd0
\kern-.05em\copy0\kern-\wd0\kern-.025em\raise.0433em\box0}
\def\D{\hbox{\hbox{${D}$}}\kern-1.9mm{\hbox{${/}$}}}
\def\kbar{\hbox{$k$}\kern-0.2true cm\hbox{$/$}}
\def\nbar{\hbox{$n$}\kern-0.23true cm\hbox{$/$}}
\def\pbar{\hbox{$p$}\kern-0.18true cm\hbox{$/$}}
\def\nhbar{\hbox{$\hat n$}\kern-0.23true cm\hbox{$/$}}

\documentstyle[epsfig,preprint,aps,floats,amssymb]{revtex}
\begin{document}
\draft
\newfont{\form}{cmss10}

\title{On the unitarity of quantum gauge theories on non-commutative spaces}
\author{A. Bassetto$^1$, L.Griguolo$^2$, G. Nardelli$^3$ and F. Vian$^1$}
\address{$^1$Dipartimento di Fisica ``G.Galilei", Via Marzolo 8, 35131
Padova, Italy\\
INFN, Sezione di Padova, Italy\\
$^2$Dipartimento di Fisica ``M. Melloni'' and
INFN, Gruppo Collegato di Parma, \\ Viale delle Scienze, 43100 Parma, Italy\\
$^3$Dipartimento di Fisica, Universit\`a di Trento,
38050 Povo (Trento), Italy \\ INFN, Gruppo Collegato di Trento, Italy}
\maketitle
\begin{abstract}
We study the perturbative unitarity of non-commutative quantum Yang-Mills
theories, extending previous investigations on scalar field theories to the
gauge case where non-locality mingles with the presence of unphysical states.
We concentrate our efforts on two different aspects of the problem. We start
by discussing the analytical structure of the vacuum polarization tensor, showing
how Cutkoski's rules and positivity of the spectral function are realized
when non-commutativity does not affect the temporal coordinate. When instead
non-commutativity involves time, we find the presence of extra
troublesome singularities on the $p_0^2$-plane that seem to invalidate the
perturbative unitarity of the theory. The existence of new tachyonic poles,
with respect to the scalar case, is also uncovered. Then we turn our attention
to a different unitarity check in the ordinary theories, namely time exponentiation of a Wilson
loop. We perform a $O(g^4)$ generalization to the (spatial)
non-commutative case of the familiar results in the usual Yang-Mills theory.
We show that exponentiation persists at $O(g^4)$ in spite of the presence of
Moyal phases reflecting non-commutativity and of the singular infrared
behaviour induced by UV/IR mixing.
\end{abstract}
\vskip 2.0truecm
DFPD 01/TH/19

\noindent
UPRF-01-012

\noindent
UTF/444

\noindent
PACS numbers: 11.15Bt, 11.15Pg, 11.15Me 

\noindent
{\it Keywords}: Unitarity, non-commutative gauge theories, 
Wilson loops.
\vskip 3.0truecm
\vfill\eject

\narrowtext

\section{Introduction}
\noindent
Recently field theories defined on non-commutative spaces have
received much attention, mostly triggered by their tight relation
with  some limiting cases of string theories \cite{cds,dh,seiwi}. These field 
theories are non-local and  non-locality has dramatic consequences on their 
basic dynamical features \cite{minwa,suski}: 
although their ultimate ``physical'' motivation is 
provided, in our opinion, by their stringy derivation, the possibility 
of exploring some specific non-local field theories in a concrete,
systematic way in search of unexpected properties is fascinating on its own.

Non-commutative field theories are explicitly constructed from the 
conventional (commutative) ones by replacing  the 
usual multiplication of fields in the Lagrangian with the
$\star$-product  of fields. 
The $\star$-product is obtained by introducing a real antisymmetric matrix 
$\theta^{\mu\nu}$ which parametrizes  non-commutativity of Minkowski 
space-time:
\beq
[x^\mu,x^\nu]=i\theta^{\mu\nu}\quad\quad\quad\quad \mu,\nu=0,..,D-1.
\eeq
The $\star$-product of two fields $\phi_1(x)$ and $\phi_2(x)$ is defined as
\beq
\label{star}
\phi_1(x)\star\phi_2(x)=\exp \lq \frac{i}2 \, \theta^{\mu\nu}\frac{\partial}{\partial x_1^\mu}
\frac{\partial}{\partial x_2^\nu}\rq \phi_1(x_1)\phi_2(x_2)|_{x_1=x_2=x}
\eeq
and  leads to terms in the action with an infinite number of derivatives of 
fields which makes the theory non-local. Then one may wonder under which
conditions the theory would still fulfill the unitarity requirements.

Unitarity is of course a central issue for the correct physical interpretation 
of a quantum field theory: as  is well known, it is granted once the 
time evolution of a system in a 
Hilbert space is driven 
by a self-adjoint operator, its Hamiltonian. In the ordinary case
the problem becomes subtle when gauge theories are concerned: the norm 
in the full 
Fock space is not positive definite, ghosts propagate as virtual 
states in physical processes and unitarity is formally recovered at the level
of the $S$-matrix 
looking at transition amplitudes between physical states (selected by the 
BRST condition). In quantum non-commutative gauge theories, where non-locality
mingles with the presence of unphysical states and with the poor definition
of $S$-matrix elements, the investigation of unitarity becomes particularly
challenging.

Unitarity of scalar field theories, in the presence of non-commutativity, has 
been discussed, in a perturbative framework, in Ref.~\cite{gomis}: the authors 
explicitly show that Cutkoski's rules are correct when $\theta^{\mu\nu}$ 
is of the ``spatial'' type, \ie $\theta^{0i}=0$. This  exactly
corresponds to the case in 
which an elegant embedding into string theory is possible: low-energy 
excitations of a $D$-brane in a magnetic background are in fact described by 
field theories with space non-commutativity \cite{seiwi}. In this limit the 
relevant description of the dynamics is in terms of massless open string 
states, while massive open string states and closed strings decouple: therefore
the full unitary string theory seems  consistently truncated to 
field-theoretical degrees of freedom, suggesting the possibility that
also related  quantum field theories are unitary.
The picture is consistent even at string-loop level as shown in
\cite{stringhi}.

On the other hand theories with $\theta^{0i}\neq 0$ have an infinite number
of time derivatives and are non-local in time: in this situation it is
not clear whether the usual framework of quantum mechanics makes sense,
in particular unitarity may be in jeopardy when the non-commutative
parameter $\theta^{\mu\nu}$ affects the time evolution, the concept of
Hamiltonian losing, in some sense, its meaning (see however \cite{kamimura}).

This fact is not surprising when observed
from the string theory point of view: $\theta^{0i}\neq 0$ is obtained in
the presence of an electric background and recent works \cite{om} have pointed
out that in the relevant low-energy limit massive open string states do
not decouple while closed strings do. The truncation of such a string theory
to its massless sector is not consistent in this case (see also
\cite{barb}).

The breakdown of unitarity in time-like scalar non-commutative
theories has been recently discussed in Ref.~\cite{alva}  and in
Ref.~\cite {mateos} in the non-relativistic case.~\footnote{The limiting
case of a light-like $\tilde{p}_\m$ has been discussed in
Ref.~\cite{ahar}. The authors show that unitarity constraints are
fulfilled also in this situation, provided the theory is formulated in 
terms of a light-front quantization.}

In this paper we study  unitarity properties of non-commutative
quantum Yang-Mills theories.

Our first effort (Sect.~II) is to
generalize and deepen the work of \cite{gomis} to the gauge theory
case:  we concentrate our
attention on
the vacuum polarization tensor $\Pi^{\mu\nu}$. We derive the complete
one-loop result in general $D=2\omega$ dimensions, using Feynman gauge.
Going to $D=4$ we recover the well-known fact that only planar diagrams
are UV-divergent while the non-planar part depends separately on two
different kinematical variables, $p^2=p_\mu p^\mu$ and $\tilde
{p}^2=\theta_{\mu\nu} p^\nu \theta^{\mu_\lambda}p_\lambda$. From the point of
view of the dispersion relations the situation is clear in the non-commutative
spatial case: the vector $\tilde p_{\mu}$ is spacelike and there is no point in
analytically continuing the variable $\tilde p^2$. We have to deal with normal 
dispersion relations in $p^2$ with the obvious presence of an extra kinematical variable.

The situation is much trickier when non-commutativity involves time. 
Then $\tilde p_{\mu}$ can also be timelike and it is not clear {\it a priori}
which variable should be analytically continued. The natural choice would be $p_0^2$
in this case, also in view of the fact that the Lorentz invariance is broken 
in such theories.
However, even with this choice, Cutkoski's rules are still invalid, as already
pointed out in Ref.~\cite{alva} for the scalar case; the presence of extra troublesome singularities
in the $p_0^2$-plane, while being an obvious sign of instability of the theory,
can hardly be explained in a perturbative context.
Moreover, both in the spatial and in the space-time non-commutative
cases, new poles appear in the one-loop resummed vector propagator for 
negative values of $p^2$ (tachyons). All these features are described
in Sect.~III.

In ordinary theories, another typical probe to check unitarity is provided 
by time exponentiation of a Wilson loop.
To be more specific, for a rectangular loop centered in the plane $(t,x)$,  
$x$ being  any spatial direction,
with sides $2T$ and $2L$,
one can show that  the Wilson loop amplitude exponentially decreases in the
large-$T$ limit, and the exponent  
  is related to the potential energy of a (very heavy)
  $q\bar q$ pair separated by a distance $2L$.
A perturbative computation of the Wilson loop has been widely used in 
commutative theories to check  unitarity, assuming 
 gauge invariance, or viceversa \cite{test}.
 
To extend this test to non-commutative theories is highly problematic, 
even in the spatial case. As a matter of fact the definition of the loop
via a non-commutative path-ordering \cite{wline,alvar,dorn}, 
has so far received a physical interpretation in the presence of
matter fields as a wave function of composite operators only in a
lattice formulation \cite{ambj}. 
Time exponentiation itself has not been proven to our knowledge,
even in the spatial case.

Sect.~IV is devoted to a perturbative ${\cal O}(g^4)$
generalization to the spatial non-commutative case of the familiar results
in the usual theory. We show that exponentiation persists at ${\cal O}(g^4)$
in spite of the phases which reflect non-commutativity.
 
Finally, in Sect.~V we draw our conclusions and discuss future developments, whereas 
technical details are deferred to  the Appendix.

\section{\bom{U(N)} Non-Commutative Yang-Mills}

\noindent
In this section we analyze the $U(N)$ Yang-Mills theory on a non-commutative space. The classical action reads 
\beq 
\label{action}
S=-\frac1{2g^2} \int d^4x\, F_{\m\n} \star  F^{\m\n}
\eeq
where the field strength $F_{\m\n}$ is given by
\beq
F_{\m\n}=\p_\m A_\n -\p_\n A_\m -i (A_\m\star A_\n - A_\n\star A_\m)
\eeq
and $A_\m$ is a $N\times N$ matrix.
The $\star$-product was defined in  Eq.~\re{star}.
The action    Eq.~\re{action} is invariant under $U(N)$
non-commutative gauge transformations 
\beq
\label{gauge}
\de_\l A_\m= \p_\m \l -i (A_\m\star\l -\l\star A_\m) \,.
\eeq
The Feynman rules for a non-Abelian non-commutative gauge theory were
worked out in \cite{armoni} and a full list is reported in the Appendix,
together with our conventions; for further investigations on quantum aspects of
non-commutative gauge theories see \cite{tutti}.  
We quantize the theory in the
Feynman-'t Hooft gauge and, in order to check unitarity and gauge
invariance at the quantum level, we consider the one-loop correction
to the gluon self energy.   One-loop
diagrams contributing to the two-point function are shown in Fig.~1.
\begin{figure}[h]
\begin{center}
\epsfxsize=15cm
\epsffile{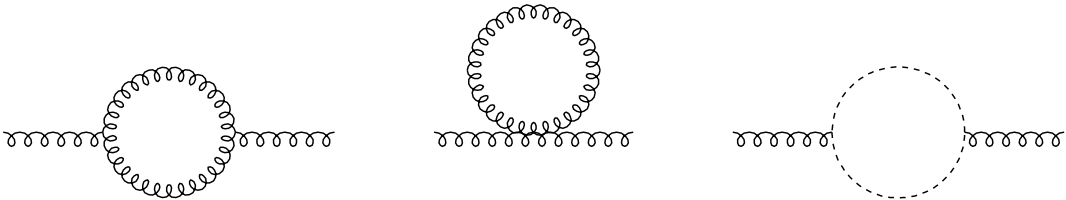}
\caption{One-loop corrections to the two-point function}
\end{center}
\end{figure}
In the ordinary 
gauge theories, tadpole diagrams vanish in
dimensional regularization  ($D=2\omega=4-2\e$).  However, this is not
true in the 
non-commutative case  and the tadpole must be included in the
computation.
By using the Feynman rules given in the Appendix the sum of the diagrams
of Fig.~1 turns out to be
\beeq
\label{two-point}
&&\Pi_{\m\n}^{AB}(p) = N (g\,\m^{2-\omega})^2 
\int \frac{d^{2\omega}q}{(2\pi)^{2\omega}}
\lq \frac{4(p^2
g_{\m\n} -p_\m p_\n)}{q^2 (p-q)^2} + 2(\omega-1) \lp \frac{(p-2q)_\m
(p-2q)_\n}{q^2 (p-q)^2}-\frac{2g_{\m\n}}{q^2}\rp \rq \no \\
&&\phantom{\Pi_{\m\n}^{AB}=N g^2 \int
\frac{d^{2\omega}q}{(2\pi)^{2\omega}}}
\times
\lp \de^{AB}-\de^{A0}\de^{B0} \cos  (\tilde{p}q)\rp  \,,
\eeeq
where $\tilde{p}^\m=\th^{\m\n}p_\n$ and $\th^{12}=-\th^{21}\equiv \th$,
all the other components vanishing.
One immediately recognizes that in  Eq.~\re{two-point}
the planar and the non-planar (\ie $\th$-dependent) contributions
can be singled out.  The term proportional to $\de^{AB}$ corresponds to
the planar diagrams \cite{minwa}, and coincides with ordinary Yang-Mills
theory with the usual $U(N)$ group factor $N \de^{AB}$. In four
dimensions this integral is divergent and produces, once regulated,
the well-known 
$1/\e$ pole. On the other hand all the novelty of
non-commutativity is concentrated in the term with $\cos
(\tilde{p}q)$, corresponding to the ultraviolet finite non-planar
contribution. Since  this term only affects the $U(1)$ propagator, in
the following we will restrict ourselves, with no loss of generality,
to the $U(1)$ case, where $\Pi_{\m\n}$ becomes
\beeq
\label{2ptabelian}
&&\Pi_{\m\n}(p)= (g\,\m^{2-\omega})^2  
\int \frac{d^{2\omega}q}{(2\pi)^{2\omega}}
\lq \frac{4(p^2
g_{\m\n} -p_\m p_\n)}{q^2 (p-q)^2} + 2(\omega-1) \lp \frac{(p-2q)_\m
(p-2q)_\n}{q^2 (p-q)^2}-\frac{2g_{\m\n}}{q^2}\rp \rq \no \\
&&\phantom{\Pi_{\m\n}=2 g^2 \int
\frac{d^{2\omega}q}{(2\pi)^{2\omega}}}
\times
\lp 1-\cos  (\tilde{p}q)\rp  \,,
\eeeq
One can easily realize that this tensor is orthogonal to $p_{\mu}$
and thereby can be written as
\beq
\label{decomp}
\Pi_{\m\n}=(g_{\m\n}p^2 - p_{\mu}p_{\nu})\Pi_1+\tilde{p}_{\m}\tilde{p}_{\n}
\Pi_2.
\eeq
In turn $\Pi_1$ contains the usual planar
part and a non-planar
one
$$\Pi_1=\Pi_1^p+\Pi_1^{np}.$$
Only $\Pi_1^p$ is ultraviolet divergent and therefore needs to be (dimensionally)
regularized; on the other hand $\Pi_1^{np}$ and $\Pi_2$ exhibit singularities
when $\tilde{p}^2=0$. 

Standard Feynman diagram techniques lead to the results
\beq
\label{polarc}
\Pi_1^p=\frac {i\, g^2}{16\pi^2}\lq {\frac {-p^2}{4\pi{\m}^2}}\rq ^{\omega -2}
\frac {6\omega -2}{2\omega -1} \,  \frac {\Gamma(2-\omega) \Gamma^2(\omega -1)}
{\Gamma(2\omega -2)},
\eeq
\beeq
\label{polarnc}
&&\Pi_1^{np}=\frac {-i\, g^2}{4\pi^2}\lq {\frac {p^2}{16\pi^2 {\m}^4 
{\tilde {p}}^2}}\rq ^{{\frac {\omega}{2}} -1}
\int_0^1 dx [x(1-x)]^{{\frac {\omega}{2}}-1}
[2-(\omega -1)(1-2x)^2]\\ \nonumber
&&
\ph{\Pi_1^{np}=\frac {-i\, g^2}{4\pi^2}\lq {\frac {p^2}{16\pi^2 {\m}^4 {\tilde {p}}^2}}\rq ^{{\frac {\omega}{2}} -1}}
\times K_{2-\omega} \lp \sqrt {x(1-x)
p^2 {\tilde p}^2}\rp
\eeeq
and
\beq
\label{polar2}
\Pi_2=\frac {i\, g^2}{4\pi^2}{[4\pi\m ^2]}^{2-\omega}
\lq {\frac {4p^2}{{\tilde {p}}
^2}}\rq ^{\frac {\omega}{2}}
\int_0^1 dx [x(1-x)]^{\frac{\omega}{2}}K_{\omega}\lp \sqrt {x(1-x)
p^2 {\tilde p}^2}\rp.
\eeq
It is easy to recognize from the above formulas  that, at fixed 
$\theta^2  p_{\perp}^2=\theta^2 (p_1^2+p_2^2)=-{\tilde p}^2>0$, 
the components of the polarization tensor are analytic functions of
the variable $p^2$, with a branch point at $p^2=0$ and a cut that can be conveniently
drawn along the positive $p^2$-axis. It is also straightforward to compute
their discontinuities
\beq
\label{disc1c}
\Delta \Pi_1^p=-{\frac {5g^2}{12\pi}} \theta (p^2) \, \theta (p^0),
\eeq
\beq
\label{disc1nc}
\Delta \Pi_1^{np}={\frac {g^2}{4\pi}}\theta (p^2) \, \theta (p^0)
\int_0^1 dx \, [1+4x(1-x)] \,J_0\lp \th \, \sqrt {x(1-x)\,
p^2 p_{\perp}^2}\rp
\eeq
and
\beq
\label{disc2}
\Delta \Pi_2=-{\frac {g^2 p^2}{\pi \, \th^2 p_{\perp}^2}}\theta (p^2)
\, \theta (p^0)
\int_0^1 dx\,
x(1-x) \,   J_2\lp \th \,\sqrt {x(1-x)\,
p^2 p_{\perp}^2}\rp,
\eeq
where we have now set $\omega=2.$ 

These discontinuities can also be computed applying Cutkoski's cutting rules
to Eq.~(\ref{2ptabelian})
\beeq
\label{disc2ptabelian}
&&\Delta \Pi_{\m\n}(p)=-g^2 
\int \frac{d^4 q}{(2\pi)^2}
\lq 4(p^2
g_{\m\n} -p_\m p_\n) + 2 (p-2q)_\m
(p-2q)_\n \rq \delta (q^2)\, \theta(q^0)  \no \\
&&\ph{\Delta \Pi_{\m\n}(p)=-g^2 }
\times\delta ((p-q)^2)\, \theta (p^0-q^0)
\lp 1-\cos  (\tilde{p}q)\rp]  \,.
\eeeq

We have checked that indeed the cutting rules hold also in this
non-commutative context, provided the non-commutative parameter
$\theta_{\m \n}$ has only spatial components. This claim has already
been presented for the scalar theory in Refs.~\cite{gomis,alva}.  In the stringy
context, this picture has its counterpart in the decoupling of massive
open and closed string states in the presence of a magnetic
background. 

The analytic properties outlined above are necessary requirements for 
fulfilling unitarity; next, positivity conditions,
crucial for the correct probabilistic interpretation of the theory,
are to be met.
When saturating the tensor \re{2ptabelian}
with a (spacelike) polarization vector $\varepsilon^{\m}$ and computing its 
discontinuity afterwards, a non-negative result is expected since, in the one-loop
case we are considering, positivity
cannot be spoilt by the extra phases affecting the vertices. 
A convenient choice is $\e_\m=\frac{{\tilde
p}_{\m}}{\sqrt{-{\tilde p}^2}}$, which gives
\beeq
\label{posit}
\Delta \lp \varepsilon \Pi \varepsilon \rp
&=&\frac{g^2 p^2}{4\pi}
\lq \frac{5}{3}-\int_0^1 dx \,[1+4x(1-x)] \,J_0\lp \th \,\sqrt {x(1-x)\,
p^2 p_{\perp}^2}\rp \right. \\ \nonumber
&+&\left. 4\int_0^1 \,dx \, x(1-x)\, J_2\lp \th \,\sqrt {x(1-x)\,
p^2 p_{\perp}^2}\rp \rq\, = \\ \nonumber
&=& \frac{g^2 p^2}{4 \pi }\lq \frac{5}{3}-3 \frac{\sin \xi}{\xi}+
4\frac{\sin \xi-\xi\cos \xi}{\xi^3}\rq \, \ge 0,\qquad \xi\equiv \frac
{\th {\sqrt {p^2 p_{\perp}^2}}}{2}.
\eeeq
The function in square brackets vanishes only at $\xi=0$
and for $\xi >0$ is indeed positive.
 
Such a positivity is strongly reminiscent of the usual behaviour
of Abelian theories defined on commutative spaces.
To clarify this issue, let us briefly recall the  general
expression for
the polarization tensor $\Pi_{\m \n}$ in those theories
\beq
\label{general}
\Pi_{\m \n}(p)=i \int d^4 x \,e^{ip(x-y)} \langle 0|T\lp {\cal J}_{\m}(x)\, {\cal 
J}_{\n}(y)\rp|
0\rangle + {\rm contact\, terms},
\eeq
the current ${\cal J}_{\m}$ being the source of the vector field and $T$ the Dyson
time-ordering operator. Current conservation
implies that $\Pi_{\m \n}$ is transverse with respect to $p_{\m}.$ 
If we saturate the tensor \re{general}
with a (spacelike) polarization vector $\varepsilon^{\m}$,
we get, for its discontinuity, the well-known result
\beq
\label{unita}
\Delta (\varepsilon\Pi \varepsilon)
\propto \sum_n |\langle 0| \varepsilon {\cal J}(0)|n \rangle|^2
\delta ^{(4)}(p-P_n),
\eeq
the vector $P_n$ being the total four-momentum of the on-shell intermediate 
state $|n\rangle.$ 
In a gauge theory some intermediate states may possess a negative norm; nevertheless
their presence is necessary to  cancel possible redundant degrees of freedom in
such a way that ``physical'' positivity is eventually recovered. 

However in the non-commutative case novel features arise: in fact, in
Eq.~\re{posit} the asymptotic value 5/3 of the planar contribution is reached after wiggling:
this feature, which is present also in scalar theories \cite{gomis,alva},
reminds of analogous quantum mechanical
effects (diffraction and interference) and is a consequence of the
non-commutativity of coordinates, which in turn entails a violation of
locality and of Lorentz invariance.  

The vanishing of the discontinuity at $p^2=0$ can be understood as a threshold effect;
nonetheless, $\xi=0$ can also mean $p_{\perp}=0$. 
The behaviour of the
theory in this limit looks peculiar and will be discussed later
on. Suffice it here to say that in the reference frame and in the
approximation we are considering, no production occurs in the plane
$p_{\perp}=0$. 
This
$non-absorbing$ phase of the theory is a novel feature compared to the
commuting case.
On the other hand $\Pi_{\m \n}$ itself is singular in
such a limit and cannot be cured.

In general non-local theories lead to amplitudes which are not 
polynomially bounded at infinity on the first Riemann sheet; 
this issue stands at the very heart
of a renormalization program and deserves thorough investigations.

In view of the non-locality of the theory, one might wonder whether amplitudes can still
be reconstructed from their imaginary parts via dispersion relations. In the instances 
we are considering the answer is affirmative, but to a certain extent.
The discontinuity $\Delta \Pi_1^{p}$ is constant and therefore $\Pi_1^{p}$ needs to
be subtracted once, as is well-known;  $\Delta \Pi_1^{np}$ vanishes at infinity
and no subtraction is {\it a priori} needed. 
The situation with $\Pi_2$ is subtler and somehow
pathological. In commutative theories subtraction constants in dispersion
relations are related to the presence of singularities in Feynman amplitudes.
The arbitrariness of such constants is in turn related to different renormalization
prescription. Thus, increasing discontinuities usually correspond to divergent
amplitudes. This is not always the case in non-commutative theories; 
as a matter of fact  
the  absolute value of $\Delta \Pi_2$ increases  at infinity in four 
dimensions.  As a consequence the
dispersion relation needs to be subtracted (once). Nevertheless we know from
Feynman diagram calculations that $\Pi_2$ is finite when
$\omega =2$  and therefore the subtraction
constant turns out to be completely determined.

One could consider the dispersion relation starting from the 
discontinuity in $2\omega$ dimensions; then, for suitable values of $\omega$,
no subtraction is necessary and, at variance with the usual case, no pole occurs in the continuation to $\omega =2$.
However this procedure is somehow extraneous to the spirit of a dispersive approach,
being possible only in a perturbative context where amplitudes can be directly
computed anyway.

In recovering Eqs.~(\ref{polarnc},\ref{polar2}) from Eqs.~(\ref{disc1nc},\ref{disc2})
respectively, one may use the Stieltjes transform
\beq
\label{stiel}
\int_0^{\infty}{\frac {dx}{x+y}} J_0\lp a\sqrt{x}\rp =2\, K_0\lp a\sqrt{y}\rp,\quad a>0,
\eeq
together with the equalities
\beq
\label{besselata1}
2\int_0^1 dx\, x(1-x)\, J_2\lp \th \,\sqrt {x(1-x)\,
p^2 p_{\perp}^2}\rp =\int_0^1 dx \,[1-6x(1-x)]\, J_0\lp \th \,\sqrt {x(1-x)\,
p^2 p_{\perp}^2}\rp
\eeq
and
\beeq
\label{besselata2}
&&2\int_0^1 dx\, x(1-x)\, K_2\lp \th \,\sqrt {-x(1-x)\,
p^2 p_{\perp}^2}\rp=- \frac{4}{\th^2 p^2 p_{\perp}^2}\\ \nonumber
&&-\int_0^1 dx \,[1-6x(1-x)]\, K_0\lp \th \,\sqrt {-x(1-x)\,
p^2 p_{\perp}^2}\rp\,.
\eeeq

It is technically difficult to control the asymptotic behaviour 
of $\Pi_1^{np}$ and $\Pi_2$
in the variable $p^2$ on the entire first Riemann sheet.
One can show that they both vanish as $(-p^2)^{-1}$ 
when $p^2\to -\,\infty$
along the real half-line, in spite of the fact that 
$\Delta \Pi_2$ diverges
when $p^2\to +\, \infty$.

The coincidence between Eqs.~(\ref{polarnc},\ref{polar2}) and the results
one obtains via dispersion relations starting from Eqs.~(\ref{disc1nc},\ref{disc2})
is a proof {\it a posteriori} that indeed their asymptotic behaviour
is compatible
with discarding the contribution to the dispersive integrals at infinity.

\section{IR properties and tachyons}

\noindent
We now comment on the infrared singularities, namely the ones 
at small ${\tilde p}^2$,
affecting $\Pi_1^{np}$ and $\Pi_2$. As is well known, they are the 
counterparts of the would-be ultraviolet singularities in the absence of the
non-commutative phase. Although quite clear from a mathematical viewpoint,
their presence is  particularly troublesome in higher order calculations where the momentum
$p$ has to be integrated over. For a proposal of using
Wilsonian methods to tame infrared singularities and to prove UV
renormalizability, in the scalar case, see \cite{gp}. 

If we recall the Dyson equation for the renormalized vector propagator
\beq
\label{dyson}
(D_{\m \n}^{(ren)})^{-1}=(D_{\m \n}^{(0)})^{-1}-\Pi_{\m \n}^{(ren)}\,,
\eeq
where $D_{\m \n}^{(0)}$ is the free propagator, it is clear that infrared
divergences of $\Pi_{\m \n}^{(ren)}$ 
cannot affect $D_{\m \n}^{(ren)}$.
Although certainly appearing order  by order in the
perturbative expansion, they might be artifact of
this expansion. On the other hand at two loops and beyond, infrared divergencies
other than iterated one-loop ones, may be generated and their explicit
form has to be taken into account in order to draw conclusions on the
finiteness of $D_{\m\n}^{(ren)}$ at small $\tilde{p}^2$.

Moreover the very possibility of 
a renormalization in a non-commutative gauge theory has not been proved beyond
one loop, to our knowledge. 

The vector propagator $D_{\m \n}^{(ren)}$ acquires the analytic structure induced 
by $\Pi_{\m \n}^{(ren)}$; it is an analytic function in the cut $p^2$-plane with
possible simple poles at negative values of $p^2$. 
However such poles 
(tachyons) would conflict with causality
and signal instability of the theory. If present, they may give rise to
spontaneous symmetry breaking after condensation.
Moreover in this context they  look dependent on the gauge choice, on the running
mass $\m$ and on the renormalization scheme. All these dependences should eventually
disappear in any realistic solution.

In particular poles of the vector propagator 
on the negative $p^2$-axis at a certain loop order 
might indicate that the perturbative approach fails
at some momentum scale and that non-perturbative effects 
may change the infrared behaviour of the theory.

In the Feynman gauge we are considering, taking the one-loop
renormalized expression for $\Pi_1$ in the $\overline{\mbox{MS}}$ scheme
\beq
\label{piunoren}
\Pi_1^{(ren)}=-\frac{i g^2}{16 \pi^2} \frac{10}3 \log \lp \frac{-p^2}{4\pi
\m^2} \rp -\frac{i g^2}{4 \pi^2} \int_0^1 dx\, [1+4x(1-x)]
\, K_0(\sqrt{x(1-x)p^2\tilde{p}^2})
\eeq
into account, Eq.~\re{dyson} can be inverted, leading to
\beq
\label{inverse} 
D_{\m\n}^{(ren)}=-\frac{i}{p^2 (1+i\,\Pi_1^{(ren)})} \lq g_{\m\n}+i\,\Pi_1^{(ren)}
\, \frac{p_\m p_\n}{p^2} - \frac{i\,\Pi_2}{p^2 (1+i\,\Pi_1^{(ren)})+i
\,\tilde{p}^2\Pi_2}\, \tilde{p}_\m \tilde{p}_\n\rq\,.
\eeq
One easily realizes from Eq.~\re{piunoren} that $\Pi_1^{(ren)}$ is
finite at $p^2=0$; therefore, after the usual subtraction in the
planar contribution, the residue on the pole at $p^2=0$ is changed
only by a finite amount and the logarithmic branch point exhibits a
``mild'' behaviour. 
We remark that in usual
Yang-Mills theory $\Pi_1^{(ren)}$ is not finite as $p^2\to 0$: the present
behaviour is a pure non-commutative effect, that can be easily understood
realizing that, in that limit, planar and non-planar contributions, in the
$U(1)$ case, conspire to cancel thanks to IR/UV duality (see \cite{ruiz}
for a discussion on this point).
At  $\tilde{p}^2=0$, the ``soft'' singularity of
$\Pi_1^{(ren)}$ and the ``hard'' one of $\Pi_2$ are completely
sterilized by the one-loop resummation, as expected.

Other possible singularities come from the vanishing of the other two
denominators. The vanishing of $(1+i\,\Pi_1^{(ren)})$ would correspond 
to  Landau poles and looks dependent on the gauge choice and on the
subtraction procedure. It could not occur for small values of the
coupling constant, nor of $\tilde{p}^2$. Much more interesting is the
possibility of a vanishing of the second denominator, which can occur
also for small $g^2$, though being a typically non-perturbative effect. If we neglect the contribution from
$\Pi_1^{(ren)}$, we have the condition
\beq
\label{suss}
\frac{g^2}{\pi^2}\int_0^1dx\, x(1-x)\, K_2\lp\sqrt{x(1-x)p^2 
{\tilde p}^2}\rp=1\,.
\eeq
In the spatial case ($\th_{12}=\th$) we have hitherto considered,
$\tilde{p}^2=-\th^2 p_{\perp}^2$ and, 
as already noticed in Refs.~\cite{minwa,ruiz}, a pole at the value 
$p^2=-\frac{\g^2}{p_{\perp}^2}$, $\g^2 \equiv
\frac{2g^2}{\pi^2 \th^2}$,
appears, for $g^2 \ll 1$ 
in the approximation of retaining only the leading term in
$K_2$. Close to this pole we find the behaviour
\beq
\label{behav}
D_{\m\n}\approx \frac{i\, \e_\m \, \e_\n}{p^2+\frac{\g^2}{
p_{\perp}^2}}\,,
\eeq
exhibiting a residue independent of $\g^2$ and of $p_{\perp}^2$ and
with the correct sign.
This solution  does not depend on the choice of the gauge parameter, as
pointed out in Ref.~\cite{ruiz}, and, at least in the one-loop approximation,
seems to support the existence of such  an instability.
Similar phenomena have also
been observed in scalar theories \cite{scali}.
It is present only in one component of the tensor, the violation of
Lorentz invariance allowing for different dispersion relations in different
projections and it is related to the would-be quadratic mass divergence
in the usual theory; as such it is expected to be gauge invariant and
disappearing in the supersymmetric case \cite{suski,valya}.
It might be interesting to check its properties in the one-loop
expression of the four-vector amplitude.

\smallskip

We now discuss the unitarity implications of when the 
non-commutative parameter
involves the time direction, namely $\theta_{03}=-\theta_{30}=\theta.$
Some results in scalar theories have been reported in Ref.~\cite{alva} together with
their interpretation in connection with string theory.

By repeating the Feynman diagram calculation one recovers the decomposition
(\ref{decomp}) and Eqs.~(\ref{polarc}-\ref{polar2}). However one has to
keep in mind that now the vector ${\tilde p}_{\mu}$ can also be timelike. 
There is therefore a delicate problem in analytically continuing 
Eqs.~(\ref{polarnc},\ref{polar2}). We can still consider the variable $p^2$
taking the relation ${\tilde p}^2=-\theta^2 (p^2+p_{\perp}^2)$ into account, 
with $p_{\perp}^2$
being kept fixed at positive values.

By looking at  Eqs.~(\ref{polarnc},\ref{polar2}) one can easily realize the presence
of two branch points at the values $p^2=0$ and $p^2=-p_{\perp}^2.$
The amplitudes are real in the gap $-p_{\perp}^2<p^2<0.$ The right-hand cut
is referred to as the usual ``physical'' cut, whereas the cut for
negative values is the non-commutative one occurring when the parameter
$\theta$ has a time component (electric case). A natural
interpretation in terms of Cutkoski's rules is available for the ``physical''
cut, whereas extra tachyonic  excitations should be invoked to
explain the presence of the other threshold \cite{alva}. 

Finally we comment on the presence of possible bound
states in this case. In the same
approximation of Eq.~(\ref{suss})
($g^2<<1$),
we have to solve the quadratic equation
\beq
\label{susst}
p^4+p^2p_{\perp}^2+\gamma^2=0,\qquad\qquad \gamma^2=
\frac{2g^2}{\pi^2\, \theta^2}\,.
\eeq
The solutions are
\beq
\label{sol}
p_{\pm}^2=\frac{1}{2}\lq -p_{\perp}^2\pm \sqrt{p_{\perp}^4-4\gamma^2}\rq\,.
\eeq
When $p_{\perp}^2> 2\gamma$,
the roots are real, in the gap between $-p_{\perp}^2$ and 0.
They represent a couple of tachyons: 
\beq
\label{tach}
D_{\m\n}^{\pm}\approx \pm \frac{i\, \e_\m \, \e_\n}{p^2-p_{\pm}^2}\,
\frac{p_{\perp}^2 \pm \sqrt{p_{\perp}^4-4\g^2}}{2\sqrt{p_{\perp}^4-4\g^2}}\,.
\eeq
In the limit $\g^2 \to 0$ ($p^2_+ \to 0\,, \quad p^2_{-}\to -p_{\perp}^2$)
\beq
\label{lim1}
D_{\m\n}^+\approx \frac{i\, \e_\m \, \e_\n}{p^2}\,,
\eeq
and the $(+)$-pole is turned into a finite correction to the free
pole, the same as in the one in the spatial case (see Eq.~\re{behav}), whereas
\beq
\label{lim2}
D_{\m\n}^-\approx -\frac{i\, \e_\m \, \e_\n}{p^2+
p_{\perp}^2}\, \frac{\g^2}{p_{\perp}^2}
\eeq
and the $(-)$-pole decouples.

When instead $p_{\perp}^2< 2\gamma$ the roots migrate to complex
conjugate values and their interpretation looks obscure.

\section{Time exponentiation of a Wilson loop as a test of unitarity}

\noindent
A fairly general non-perturbative test of unitarity in the usual
commutative case is provided by the time exponentiation of a Wilson
loop \cite{bass1,bass2}.  To be more specific, one considers a
rectangular Wilson loop, centered in the origin of the $(t,x)$-plane
with sides of length $2T,2L$, respectively.  One can show that, in the
large-$T$ limit, its expression coincides (apart from a trivial
threshold factor) with the vacuum--to--vacuum overlap amplitude of two
$q\bar q$ strings at times $-T$ and $T$ respectively.  The (very
heavy) quarks are kept at a fixed finite distance $2L$.

By expanding on the complete set of energy eigenfunctions, after time
translations, the loop acquires the expression (for Euclidean time)
\begin{equation}
\label{expo}
{\cal W}(T,L)=\exp(-2{\cal E}_0 T)\int_{{\cal E}_0}^\infty d{\cal E} 
\,\rho({\cal E},L)\,\exp[-2T({\cal E}-{\cal E}_0)],
\end{equation}
${\cal E}_0(L)$ being the ground state energy of the $q\bar q$ system.
The spectral density $\rho$ is a positive measure, as a consequence of
unitarity.

The above equation implies an exponential decrease of the Wilson loop
with time. If ${\cal E}_0(L)$ increases linearly at large $L$ ( ${\cal E}_0(L)
\simeq 2\sigma L$), an area-law behaviour is obtained. The $q\bar q$-potential
confines with a string tension $\sigma$.

Although the above arguments are non-perturbative, a perturbative analysis
of the Wilson loop is interesting on its own. 
The perturbative Wilson loop  manifests a feature, which is known in the 
literature as the non-Abelian cancellation of the ${\cal O}(g^4)$ $T^2$-terms.
This property is compatible, at ${\cal O}(g^4)$, with its perturbative
exponentiation  in the large-$T$ limit.

In this Section we shall scrutinize  the perturbative 
${\cal O}(g^4)$  large-$T$ behaviour 
of the Wilson loop in the spatial non-commutative
case and compare  the results  with those of the corresponding
commutative theory, although no exponentiation property has been
proven to our knowledge nor any connection with a possible
$q\bar{q}$-potential established, at least in a continuum formulation. 

In the non-commutative case the  Wilson loop can be defined by means of the
Moyal product as \cite{wline,alvar}  
\beq
\label{wloop}
\WW[C]=\int \DD A \, e^{\,iS[A]} \int d^4x\,\tr P_{\star} \exp \lp
i\int_C A_\m 
(x+\xi(s))\, d\xi^\m(s)\rp \,,
\eeq
where $C$ is a closed contour in non-commutative space-time
parametrized by $\xi(s)$, with $0 \leq s \leq 1$, and $P_\star$
denotes non-commutative path ordering along $x(s)$ from right to left
with respect to increasing $s$ of $\star$-products of functions.
Gauge invariance requires integration over coordinates, which is
trivially realized when considering  vacuum averages \cite{dorn}.

We consider the closed path $C$ parametrized by the following four
segments $\g_i$
\beq\label{contour}
\begin{array}{lclcl}
\g_1 &:& \g_1^\m (s)&=&(-sT, L),\\
\g_2 &:& \g_2^\m (s)&=&(-T, -sL),\\
\g_3 &:& \g_3^\m (s)&=&(sT, -L),\\          
\g_4 &:& \g_4^\m (s)&=&(T, sL),        
\qquad \qquad -1\leq s\leq1\,,
\end{array}
\eeq
describing a (counterclockwise-oriented) rectangle centered at the
origin of the plane ($x_0$, $x_3$), with length sides ($2T$, $2L$),
respectively. The perturbative expansion of $\WW [C]= \WW   (T,L)$,
expressed by Eq.~\re{wloop}, reads (for a $U(1)$ non-commutative theory)
\beeq \label{loopert}
\WW   (T,L)&=&
\langle 0 |  {\cal T} (P_\star e^{ig\int_C A_\m
dx^\m})| 0\rangle  \\
&=&
\sum_{n=0}^\infty (ig)^n \int_{-1}^1 ds_1 \ldots
\int_{s_{n-1}}^1 ds_n 
\, \dot{x}^{\m_1}\ldots\dot{x}^{\m_n}\no\\
&&\ph{
\sum_{n=0}^\infty (ig)^n } 
\langle 0\left| {\cal T}\lq  A_{\m_1}(x(s_1))
\star\ldots\star    A_{\m_n}(x(s_n))\rq \right|0\rangle \no
\eeeq
and it is
easily shown to be an even power series in $g$, so that we can write
\beq
\label{wpert}
\WW   (T,L)= 1 +g^2 \WW_2+g^4\WW_4+\OO(g^6)\,.
\eeq
Through an explicit evaluation one is convinced that the function
$\WW_2$ in Eq.~\re{wpert} is reproduced by the single-exchange diagram 
(Fig.~2), which is exactly as in the ordinary $U(1)$ theory.
\begin{figure}[h]
\begin{center}
\epsfxsize=6cm
\epsffile{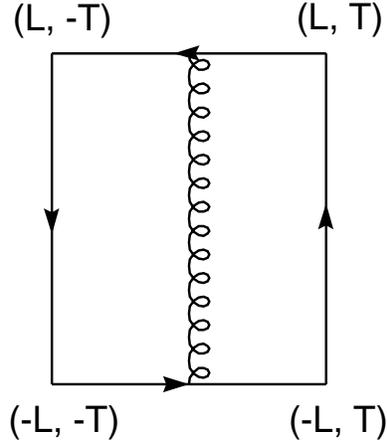}
\caption{Single exchange}
\end{center}
\end{figure}

On the other hand the diagrams contributing to $\WW_4$ can be grouped
into three distinct families:
\begin{itemize}
\item those with a double vector exchange in which propagators either do not
cross ($\WW_{nc}$) or cross ($\WW_{c}$);
\item
those involving a vertex ($\WW_s$);
\item 
those with a one-loop self-energy insertion in the free propagator
($\WW_b$).
\end{itemize}

In the large-$T$ limit the leading contribution to $\WW_{nc}$ is
depicted in Fig.~3 and is given by
\beeq
\label{noncross}
&&\WW_{nc}(T,L)=-\frac{g^4}{(2\pi)^4}
\int \frac{d^4 p}{p^2} \, \frac{d^4q}{q^2} \int_{-T}^T dx 
\int_{x}^T ds \int_{T}^{-T} dw \int_{w}^{-T} dy \, \,e^{i\lq p_0 (x
-y)+q_0 (s-w)\rq}    \, e^{2iL(p_3+q_3)} \no \\
&&\ph{\WW_{nc}(T,L)}
\approx -\frac{g^4 T^2}{\pi^2} \lp  \int d^3p
\frac{e^{2ip_3L}}{\vec{p\,}^2}\rp^2   = -g^4 \pi^2 \lp\frac{T}L\rp^2\,.
\eeeq
We derived such an expression by first rescaling the variables $x$, $y$, $s$, $w$
and then neglecting terms like ${(\frac{p_0}{T})}^2$, ${(\frac{q_0}{T})}^2$ 
with respect to ${\vec{p\,}^2}$ and ${\vec{q\,}^2}$
\cite{bassold}. 
\begin{figure}[h]
\begin{center}
\epsfxsize=6cm
\epsffile{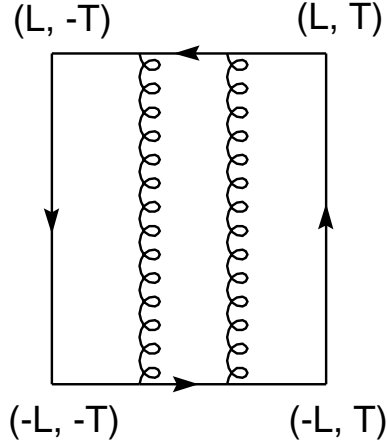}
\caption{Non-crossed vector exchange}
\end{center}
\end{figure}

Our goal is to prove that Eq.~\re{noncross} represents the only
leading contribution to $\WW_4$ also in the non-commutative
case. Hence we proceed examining  $\WW_c$. The potentially $\OO (T^2)$ 
crossed diagrams are those shown in Fig.~4  together with their
symmetric under the exchange $\g_1 \to \g_3$. 
\begin{figure}[h]
\begin{center}
\epsfxsize=15cm
\epsffile{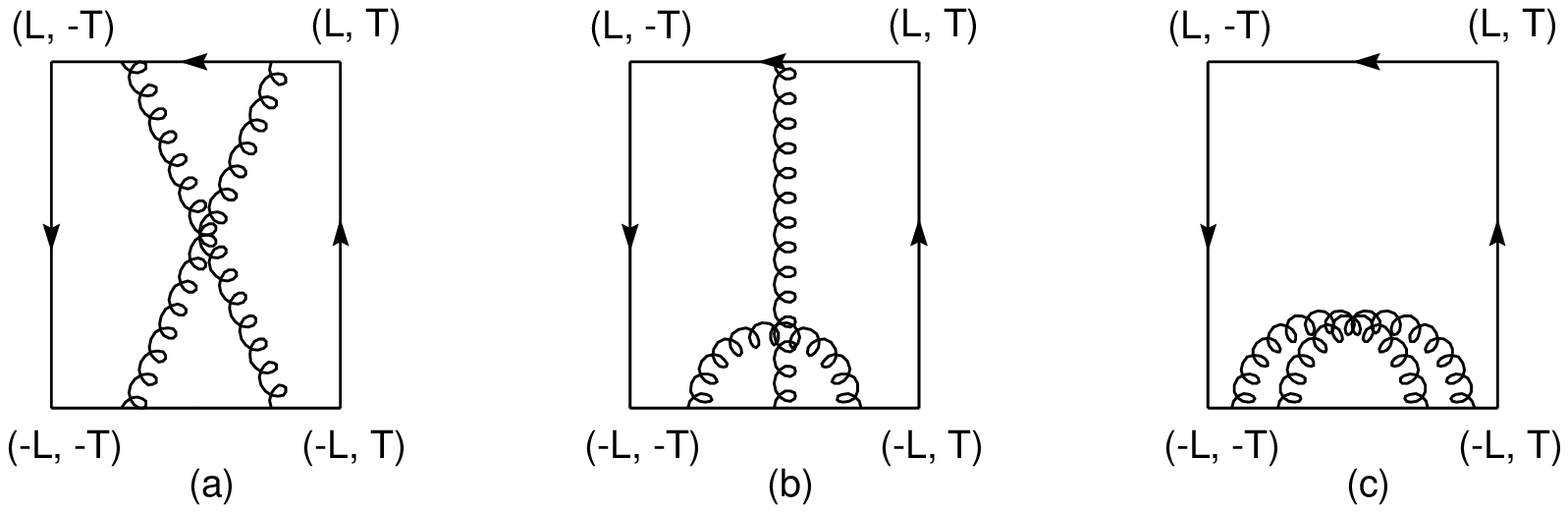}
\caption{Dominant crossed  vector exchanges}
\end{center}
\end{figure}
As a representative, let 
us consider the graph depicted in Fig.~4(a), having  the expression
\beeq
\label{cross}
\WW_{c}(T,L)&=&-\frac{g^4}{(2\pi)^4}
\int \frac{d^4 p}{p^2}\, \frac{d^4 q}{q^2} \int_{-T}^T dx 
\int_x^T ds \int_{T}^{-T} dy \int_{y}^{-T} dw \no\\
&&\ph{\frac{g^4}{(2\pi)^4}\,\,}
\times   \,
e^{i\tilde{p}q}
\, e^{i\lq p_0 (x
-y)+q_0 (s-w)\rq}    \, e^{2iL(p_3+q_3)} \\
&\approx &
\frac{g^4 T^2}{4\pi^4}\int dp_0 \,\frac{dq_0}{q_0^2} 
\lq e^{iq_0}\frac{\sin (p_0)}{p_0}-\frac{\sin (p_0+q_0)}{p_0+q_0}\rq^2  
\int d^3p\,d^3q \,e^{2i(p_3+q_3)L}\,
\frac{e^{i\tilde{p}q}}{\vec{p\,}^2\vec{q\,}^2}\,.\no
\eeeq
Integrations over $q_0$ and $p_0$ are easily proven to produce a
vanishing result. Similarly one can treat integrals arising from
graphs in Figs.~4(b), (c).

The quantity $\WW_s$ comes from ``spider'' diagrams, namely those
containing the triple vector vertex. 
\begin{figure}[h]
\begin{center}
\epsfxsize=6cm
\epsffile{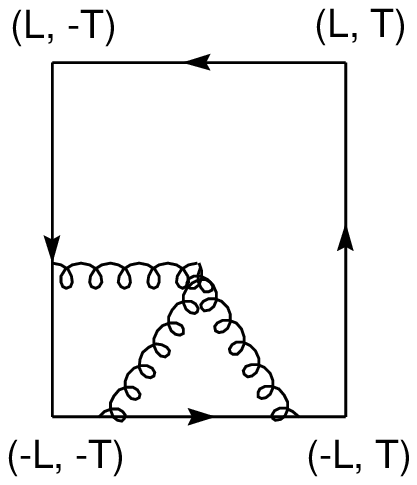}
\caption{The triple vertex diagram $S_{233}$}
\end{center}
\end{figure}
It can be straightforwardly
checked  that spider diagrams are at most $\OO(T^0)$. By denoting by
$S_{ijk}$ the contribution of the diagram in which the vectors are 
attached to the lines $\g_i$, $\g_j$, $\g_k$, respectively, one has
for instance (see Fig.~5)
\beeq
\label{spider}
S_{233}&=&\frac{g^4}{(2\pi)^4}
\int d^4 p\, d^4 q \int_{-T}^T dx 
\int_x^T dy \int_{L}^{-L} ds \,V_{003}(p,q,k;\th) \no\\
&&\ph{\frac{g^4}{(2\pi)^4}\,\,}
\times   \,
e^{\frac{i}2\tilde{p}q}
\, e^{i  p_0 x} \,e^{i  q_0 y} \,e^{-i  k_0 T} \,e^{-i  k_3 s}
\, e^{iL(p_3+q_3)} \,,
\eeeq
where
$$
V_{003}(p,q,k;\th)=2 \,\frac{(p_3-q_3) \sin\lp\frac{\tilde{p}q}2\rp
}{p^2\, q^2\, k^2}
$$
and $k=-p-q$.
After integrating out the geometric variables $x$ and $y$ and  adopting the
approximation introduced in \re{noncross}, Eq.~\re{spider} reduces to
\beeq
\label{spider1}
S_{233}&\approx&
\frac{i \,g^4}{4\pi^4}
\int dp_0 \,dq_0 
\int_L^{-L}ds  \int \frac{d^3p\,d^3q}{\vec{p}\,^2\,\vec{q}\,^2\, (\vec{p}+\vec{q})\,^2} \,
\frac{e^{i(p_0+q_0)}}{q_0} \,
\lq e^{iq_0}\frac{\sin (p_0)}{p_0}-\frac{\sin (p_0+q_0)}{p_0+q_0}\rq \no\\
&&\ph{\frac{i \,g^4}{4\pi^4}\,\,} \times e^{\frac{i}2\tilde{p}q}\,
e^{i(p_3+q_3)L}\,e^{-i  k_3 s}\,
(p_3-q_3) \,\sin\lp\frac{\tilde{p}q}2\rp
\eeeq
which is manifestly $\OO(T^0)$.

We now turn to the calculation of $\WW_b$ in the large-$T$ limit,
namely of the diagrams with a single vector exchange and a self-energy 
correction $\OO(g^2)$ (``bubble'' diagrams). We indicate with $B_{ij}$ 
the diagram in which the propagator, given by
Eqs.~\re{polarc}-\re{polar2}, connects the sides $\g_i$, $\g_j$.
Among all $B_{ij}$'s, in the limit $T\to \infty$, leading
contributions are produced by $B_{11}=B_{33}$ and $B_{13}$; however they
turn out to be only $\OO(T)$. Those diagrams are represented in Fig.~6.
\begin{figure}[h]
\begin{center}
\epsfxsize=12cm
\epsffile{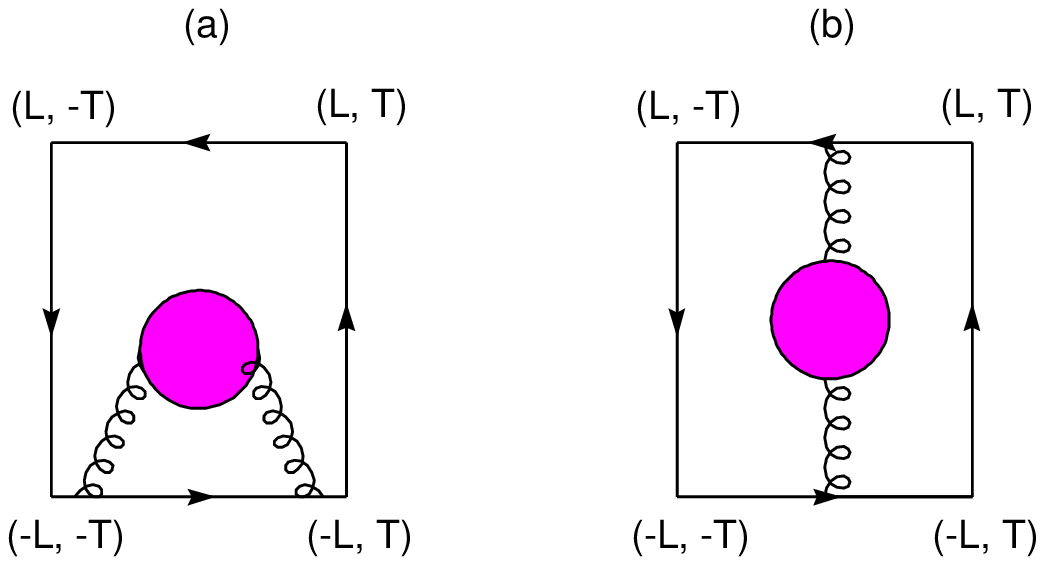}
\caption{Dominant diagrams with a one-loop self-energy insertion:
$B_{33}$ in (a) and $B_{13}$ in (b)}
\end{center}
\end{figure}
Nonetheless there are subtleties due to the fact that the planar part
of the self-energy is UV divergent; henceforth, when inserted in the
Wilson loop, the latter needs being regularized. The same problem
clearly occurs also in the ordinary case and is usually dealt with by
considering the analytic continuation of the self-energy in the
complex $\omega$-plane and by performing the limit $T\to \infty$ while
keeping $\omega\neq 2$ \cite{bassold}.
Once this point has been made clear, $B_{33}$ reads
\beq
\label{bubble}
B_{33}=\frac{(g\,\m^{2-\omega})^2}{(2\pi)^{2\omega}}
\int \frac{d^{2\omega} p}{p^4}\int_{-T}^T dx 
\int_{x}^T dy 
\, e^{i  p_0 (x-y)} \, (p^2-p_0^2)\, \Pi_1\,.
\eeq 
After integrating out $x$ and $y$, one can show that,
in the large-$T$ limit, $B_{33}$ does not increase faster than $T$.

In the same philosophy of Ref.~\cite{bassold}, such a contribution,
divergent as $\omega \to 2$ but subleading in $T$, is
discarded. The same large-$T$ behaviour is exhibited by  
$B_{13}$, the only essential difference being the phase factor $\exp
(2ip_3L)$, independent of the geometrical variables, in the integrand.

With the choice of non-commutative parameter we have hitherto
considered ($\th_{12}\neq 0$), the peculiar extra structure encoded in
$\Pi_2$ has not been probed. A more intriguing situation occurs if
we scrutinize the case $\th_{23}\neq 0$. In this case the UV-IR mixing 
phenomenon, typical of a non-commutative theory, affects the
$\OO(g^4)$ loop calculation. As a matter of fact, $\Pi_2$ diverges as  
$(p_\perp^2)^{-2}$ in the IR, which is the counterpart of the would-be
quadratic mass singularity coming from a tadpole. If we again adopt
the philosophy of keeping $\omega\neq 2$, the singularity is not exposed
and we recover a sub-leading behaviour in the large
$T$-limit. However, there is an additional problem with respect to the 
previous case, in as much as we have to invoke continuation in $\omega$ in 
order to regularize both the UV and the IR behaviours.

One could assume a different attitude, namely consider a one-loop
resummed vector propagator, according to Eq.~\re{inverse}, keeping the
dependence on the renormalization procedure, which should hopefully
be irrelevant to the final result. 
Then, of course, IR singularities get sterilized; nonetheless, this
choice should be consistently performed in all the diagrams we have
considered and would imply an infinite partial resummation of
perturbative diagrams, leading thereby to a result beyond
$\OO(g^4)$. In addition, one would face the troublesome problem of the 
appearance of the tachyonic pole we have described in the previous
section, whose extra contribution could hardly be interpreted.

As a final remark, we stress that our treatment of the Wilson loop
applies to a non-commutative theory of a ``magnetic'' type. In the
``electric'' case, leaving aside the difficulties related to the
interpretation of such a Wilson loop in this situation, 
its large-$T$ behaviour involves the non-commutative parameter via the 
Moyal phase. Not only one encounters considerable technical
difficulties in performing such a computation, but also different  
scaling limits lead to different outcomes and
all of them still call for a sensible explanation.

\section{Conclusions}

\noindent
We  discussed the quantum consistency of non-commutative gauge
theories by investigating their unitarity properties at perturbative level. In the
first part of the paper we  extended the work of \cite{gomis,alva} to gauge
theories, studying the one-loop level  analytical structure of the vacuum
polarization tensor both for pure spatial (magnetic) and space-time
(electric) non-commutative parameter. The general feature is a
violation of Lorentz covariance through  the appearance
in the amplitudes of a new kinematical variable $\tilde
p^2=\theta_\mu^\nu\theta^\mu_\lambda p_\nu p^\lambda$; dispersion
relations are strongly affected by its presence. In the magnetic case we
 found that Cutkoski's rules are satisfied by considering only physical
branch cuts: the positivity of spectral densities related to transverse
polarizations is checked, although being realized through an oscillating
behaviour, and the possibility of recovering the full Feynman amplitude
through its imaginary part, in spite of the non-locality of the theory,
was carefully discussed. On the other hand,
in the electric case we saw the appearance of extra singularities on
the $p_0^2$-plane: a threshold starting at $p^2=-p_\perp^2$, with
a non-positive definite discontinuity, suggests the presence of tachyonic
excitations carrying negative probabilities. Perturbative unitarity is
therefore lost. 

Next, by resumming the one-loop result, new poles for
physical polarizations come into play: it is well known \cite{minwa,suski} that,
thanks to the IR/UV mixing, it is possible to obtain isolate poles, for
energy well below the (usual) Landau singularity,  \ie for small
momenta. In the pure spatial case a tachyon is found
with  positive probability,  signalling,
anyway, a perturbative instability. For a space-time non-commutative
parameter the situation is much more exotic: two tachyonic poles appear for
$-p_\perp^2<p^2<0$ at perturbative (\ie $\OO (g^2)$) momenta, with different
positivity properties. The ghostly one decouples as $g^2\to 0$ while the
other is turned, in the same limit, into a correction of the free pole. Above
a certain value of the coupling both poles migrate into the complex plane.

In the second part of the paper, we proposed an extension of the usual
test of time exponentiation of a Wilson loop to the non-commutative
gauge theories: the definition of the loop through non-commutative
path-ordering makes its physical interpretation not straightforward
even in the pure spatial case.  We nevertheless showed in the magnetic
case that exponentiation persists at $\OO(g^4)$, in spite of the
presence of Moyal phases and of the appearance of new infrared
singularities. The approximations we employed in the computation are
likely to be justified in the spatial case while for space-time
non-commutativity they are likely to be invalid, posing a difficulty
of principle to the investigation.

Our results fit well with the common wisdom derived from the stringy
picture, relating the magnetic case to a good field theoretical limit;
the non-unitarity of the electric case has to be ascribed, instead, to
the impossibility of decoupling massive open-string states from the
light degrees of freedom. Many questions remain, nevertheless, to be
answered. The first one concerns the presence of a tachyonic pole
in the ``magnetic'' theory, that seems to signal an instability of the
perturbative vacuum. It would be very interesting to study its effect
on the four-point function, where unitarity poses strong constraints
through the presence of crossed channels.
The very same
singularity seems to be an obstacle for the renormalizability program:
in particular the recent proposal \cite{gp} to define an IR safe
perturbation theory through resummation appears in conflict with the
presence of a tachyonic pole. A more careful investigation of the
vacuum properties of non-commutative gauge theories is probably
needed. On the other hand it would be important to better understand 
if the Wilson loop test is fully justified in the non-commutative
context. Wilson lines have a natural interpretation when coming from
string theory \cite{wline}, and therefore their relation with unitarity is
likely to be simpler in that context.  

All these issues are
currently under investigations.

\section{Appendix A - Feynman rules for the non-commutative $U(N)$ Yang-Mills 
theory}

\noindent
The non-commutative Yang-Mills action in the 't Hooft-Feynman gauge
including ghosts takes the following form
\beq
S= \int d^4 x\  \tr \ \left (-{1\over 2 g^2 } F^{\mu \nu} \star F  _{\mu \nu}
+ (\partial ^\mu A _\mu)^2 - 
 \bar c \star \partial ^\mu D_\mu c 
+ \partial ^\mu D_\mu c \star \bar c \right )\,.
\eeq  
In the Feynman-'t Hooft gauge, the Feynman rules are  
\unitlength 1mm
\beeq
\begin{picture}(30,8)
\put(-10,-11){
\parbox{4.2cm}{
\epsfig{figure=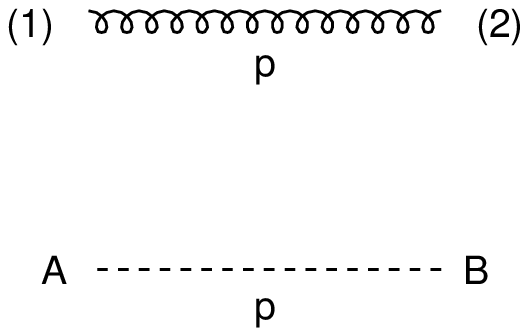,width=4.2cm}}
}
\end{picture}
&&\qquad \,\, -\frac{i}{p^2} \, \de^{AB} g^{\m\n} \no\\
&&\no\\
&&\qquad \, \, \frac{i}{p^2} \, \de^{AB}\no \\
%&&\no\\
\begin{picture}(30,8)
\put(-10,-50){
\parbox{4.2cm}{
\epsfig{figure=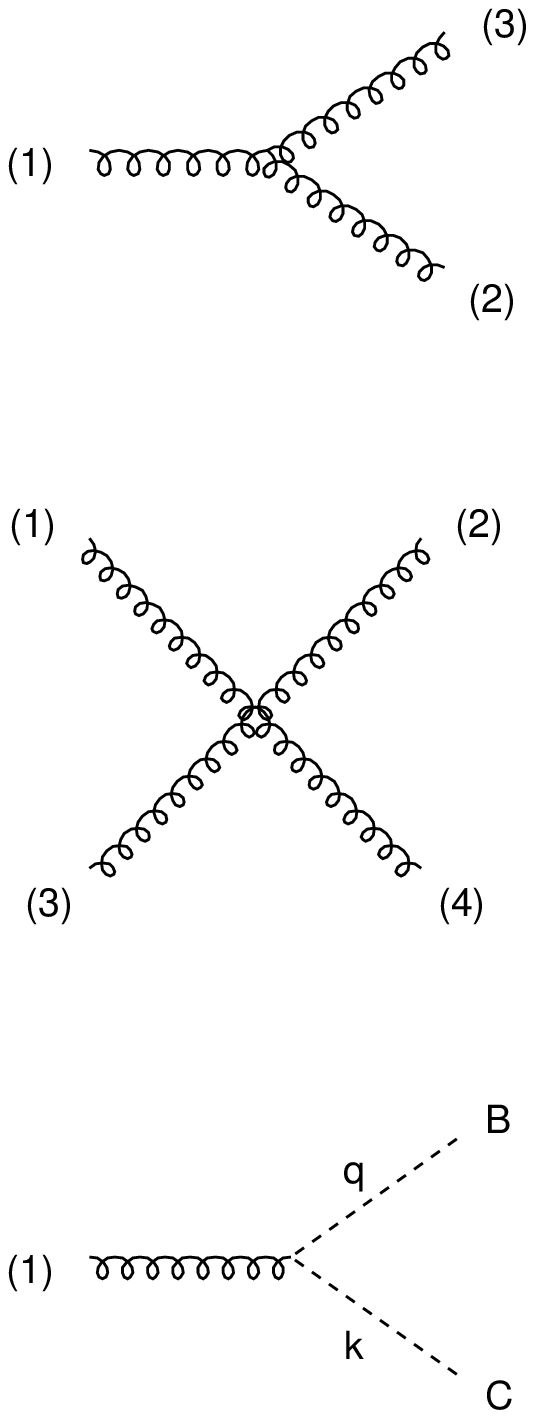,width=4.2cm}}
}
\end{picture}
&&\qquad \no\\
&&\qquad 2g \lp -i \cos\lp\frac{\tilde{p}q}2\rp \,
\tr [t^A,t^B]t^C +
\sin\lp\frac{\tilde{p}q}2\rp \,
\tr \{t^A,t^B\}t^C \rp \no \\
&& \qquad \quad \,\,\times \lq (k-p)^\n g^{\m\r} + (p-q)^\r g^{\m\n}
+(q-k)^\m g^{\n\r}\rq \no\\
&&\no\\
&&\no\\
&& \qquad -2ig^2 \tr \Biggl[ \Biggl( -i 
\cos\lp\frac{\tilde{p}q}2\rp \,
[t^A,t^B] +
\sin\lp\frac{\tilde{p}q}2\rp \,
\{t^A,t^B\} \Biggr) \Biggr. \no \\
&&\qquad \ph{-2ig^2 \,\,\, } \times  \Biggl. \Biggl( -i 
\cos\lp\frac{\tilde{k}l}2\rp \,
[t^C,t^D] +
\sin\lp\frac{\tilde{k}l}2\rp \,
\{t^C,t^D\} \Biggr) \Biggr] \no \\
&& \qquad \ph{ -2ig^2 \,\,\, } \times \lp g^{\m\r} g^{\n\s} - g^{\m\s} g^{\n\r}\rp +
(1324)+(1423)\no \\
&&\no \\
&&\no \\
&& \qquad -2g \, q^\m \lp -i 
\cos\lp\frac{\tilde{p}q}2\rp \,
\tr [t^A,t^B]t^C +
\sin\lp\frac{\tilde{p}q}2\rp \,
\tr \{t^A,t^B\}t^C \rp \,,\no\\
&&\no\\
&&\no
\eeeq
where wavy and dotted lines denote gluons and ghosts, respectively,
capital letters $U(N)$ indices, small letters momenta. Finally we set
$(1)\equiv (A,\,p,\, \m )$, 
$(2)\equiv (B,\,q,\, \n )$, 
$(3)\equiv (C,\,k,\, \r )$, 
$(4)\equiv (D,\,l,\, \s )$. 

We use hermitian gauge-group generators $t^A$ with the normalization
$\tr (t^A\,t^B)=\half \de^{AB}$.

\end{document}